\documentstyle[graphicx,amssymb]{./mn}

\newif\ifAMStwofonts

\def\mnras{MNRAS}
\def\apj{ApJ}
\def\apjl{ApJL}
\def\apjs{ApJS}
\def\aap{A\&A}
\def\aaps{A\&AS}
\def\aj{AJ}
\def\nat{Nat}

\def\pasp{PASP}

\renewcommand{\dagger}{\bullet}
\renewcommand{\S}{\star}
\renewcommand{\P}{\bigtriangleup}

\ifoldfss
  \ifCUPmtlplainloaded \else
    \NewTextAlphabet{textbfit} {cmbxti10} {}
    \NewTextAlphabet{textbfss} {cmssbx10} {}
    \NewMathAlphabet{mathbfit} {cmbxti10} {} 
    \NewMathAlphabet{mathbfss} {cmssbx10} {} 
  \fi
  \ifAMStwofonts
    \ifCUPmtlplainloaded \else
      \NewSymbolFont{upmath} {eurm10}
      \NewSymbolFont{AMSa} {msam10}
      \NewMathSymbol{\upi}     {0}{upmath}{19}
      \NewMathSymbol{\umu}     {0}{upmath}{16}
      \NewMathSymbol{\upartial}{0}{upmath}{40}
      \NewMathSymbol{\leqslant}{3}{AMSa}{36}
      \NewMathSymbol{\geqslant}{3}{AMSa}{3E}

    \fi
  \fi
\fi 

\ifnfssone
  \newmathalphabet{\mathit}
  \addtoversion{normal}{\mathit}{cmr}{m}{it}
  \addtoversion{bold}{\mathit}{cmr}{bx}{it}
  \newmathalphabet{\mathbfit} 
  \addtoversion{normal}{\mathbfit}{cmr}{bx}{it}
  \addtoversion{bold}{\mathbfit}{cmr}{bx}{it}
  \newmathalphabet{\mathbfss} 
  \addtoversion{normal}{\mathbfss}{cmss}{bx}{n}
  \addtoversion{bold}{\mathbfss}{cmss}{bx}{n}
  \ifAMStwofonts
    \ifCUPmtlplainloaded \else
      \UseAMStwoboldmath
      \makeatletter
      \new@mathgroup\upmath@group
      \define@mathgroup\mv@normal\upmath@group{eur}{m}{n}
      \define@mathgroup\mv@bold\upmath@group{eur}{b}{n}
      \edef\UPM{\hexnumber\upmath@group}
      \new@mathgroup\amsa@group
      \define@mathgroup\mv@normal\amsa@group{msa}{m}{n}
      \define@mathgroup\mv@bold\amsa@group{msa}{m}{n}
      \edef\AMSa{\hexnumber\amsa@group}
      \makeatother
      \mathchardef\upi="0\UPM19
      \mathchardef\umu="0\UPM16
      \mathchardef\upartial="0\UPM40
      \mathchardef\leqslant="3\AMSa36
      \mathchardef\geqslant="3\AMSa3E
    \fi
  \fi
\fi 

\ifnfsstwo
  \DeclareMathAlphabet{\mathbfit}{OT1}{cmr}{bx}{it}
  \SetMathAlphabet\mathbfit{bold}{OT1}{cmr}{bx}{it}
  \DeclareMathAlphabet{\mathbfss}{OT1}{cmss}{bx}{n}
  \SetMathAlphabet\mathbfss{bold}{OT1}{cmss}{bx}{n}
  \ifAMStwofonts
    \ifCUPmtlplainloaded \else
      \DeclareSymbolFont{UPM}{U}{eur}{m}{n}
      \SetSymbolFont{UPM}{bold}{U}{eur}{b}{n}
      \DeclareSymbolFont{AMSa}{U}{msa}{m}{n}
      \DeclareMathSymbol{\upi}{0}{UPM}{"19}
      \DeclareMathSymbol{\umu}{0}{UPM}{"16}
      \DeclareMathSymbol{\upartial}{0}{UPM}{"40}
      \DeclareMathSymbol{\leqslant}{3}{AMSa}{"36}
      \DeclareMathSymbol{\geqslant}{3}{AMSa}{"3E}
    \fi
  \fi
\fi 

\ifCUPmtlplainloaded \else
  \ifAMStwofonts \else 
    \def\upi{\pi}
    \def\umu{\mu}
    \def\upartial{\partial}
  \fi
\fi

\title[Ground-state CO emission in the quasar 3C318]{Ground-state $^{12}$CO emission and a resolved jet at 115 GHz (rest-frame) in the radio loud quasar 3C318}
\author[Heywood et al.]{Ian Heywood$^{1,2}$\thanks{{\tt ianh@astro.ox.ac.uk}}, Alejo Mart\'{i}nez-Sansigre$^{3}$\thanks{Current address: \emph{CGG Veritas, Av. Pres. Wilson 231, Suite 1501,  Centro, Rio de Janeiro, RJ, Brazil}}, Chris J. Willott$^{4}$ and Steve Rawlings$^{1}$\\
$^{1}$Astrophysics, Department of Physics, University of Oxford, Keble Road, Oxford, OX1 3RH, UK\\
 $^{2}$Department of Physics and Electronics, Rhodes University, PO Box 94, Grahamstown, 6140, South Africa\\
$^{3}$Institute of Cosmology and Gravitation, University of Portsmouth, Dennis Sciama Building, Burnaby Road, Portsmouth, PO1 3FX, UK\\
$^{4}$Canadian Astronomy Data Centre, National Research Council Canada, 5071 West Saanich Rd, Victoria, BC V9E 2E7, Canada}
	
\begin{document}

\date{Accepted 2013 August 12.  Received 2013 August 11; in original form 2012 June 1}

\pagerange{\pageref{firstpage}--\pageref{lastpage}} \pubyear{20XX}

\maketitle

\label{firstpage}

\begin{abstract}

An analysis of 44 GHz VLA observations of the $z$~=~1.574 radio-loud quasar 3C318 has revealed emission from the redshifted $J$~=~1~$\rightarrow$~0 transition of the CO molecule and spatially resolved the 6.3 kpc radio jet associated with the quasar at 115 GHz rest-frame. The continuum-subtracted line emitter is spatially offset from the quasar nucleus by 0.33" (2.82 kpc in projection). This spatial offset has a significance of $>$8-sigma and, together with a previously published -400~km~s$^{-1}$ velocity offset measured in the $J$~=~2~$\rightarrow$~1 CO line relative to the systemic redshift of the quasar, rules out a circumnuclear starburst or molecular gas ring and suggests that the quasar host galaxy is either undergoing a major merger with a gas-rich galaxy or is otherwise a highly disrupted system. If the merger scenario is correct then the event may be in its early stages, acting as the trigger for both the young radio jets in the quasar and a starburst in the merging galaxy. The total molecular gas mass in the spatially offset line emitter as measured from the ground-state CO line M$_{H_{2}}$~=~3.7~($\pm$0.4)~$\times$~10$^{10}$ ($\alpha_{CO}/0.8$) M$_{\odot}$. Assuming that the line-emitter can be modelled as a rotating disk, an inclination-dependent upper limit is derived for its dynamical mass $M_{dyn}$~sin$^{2}$($i$)~$<$~3.2~$\times$~10$^{9}$~M$_{\odot}$, suggesting that for M$_{H_{2}}$ to remain less than M$_{dyn}$ the inclination angle must be $i$~$<$~16$^{\circ}$. The far infrared and CO luminosities of 246 extragalactic systems are collated from the literature for comparison. The high molecular gas content of 3C318 is consistent with that of the general population of high redshift quasars and sub-millimetre galaxies. 


\end{abstract}


\section{Introduction}
%

Two pieces of observational evidence suggesting that the growth of the stellar population in a galaxy is linked to that of the central supermassive black hole are: (i) the rise of both the cosmic star formation rate density (i.e.~the rate at which the Universe is forming stars per unit volume; Madau et al., 1998) and rate of black hole growth (Dunlop \& Peacock, 1990) over cosmic time, peaking at $z$~$\sim$~1--3 and subsequently declining to the present day; (ii) the tight correlation between the mass of supermassive black holes and the mass of their associated galaxy spheroids (Merritt \& Ferrarese, 2001). 

The deployment of the Sub-millimetre Common User Bolometer Array (SCUBA) in 1996 led to the discovery that a significant fraction of star formation in the high redshift Universe was occurring in systems containing large quantities of dust (Smail et al., 1997), which was both obscuring the starlight of the host galaxy at optical wavelengths and simultaneously re-radiating it at far infrared (FIR) wavelengths. Dubbed `sub-millimetre galaxies', these systems exhibit prodigious star formation rates ($\gtrsim$10$^{3}$~M$_{\odot}$~yr$^{-1}$) and are likely to represent the most massive elliptical galaxies undergoing formation (Hainline et al., 2011). Given the observationally-implied coeval growth of stellar bulges and black holes, and the prevalence of supermassive black holes at the centres of most galaxies (Magorrian et al., 1998), one might expect sub-mm galaxies to also display high levels of black hole growth. However X-ray observations, probing black hole accretion and largely unaffected by the dust obscuration, show that although the majority of sub-mm galaxies exhibit X-ray emission consistent with continuous low-level black hole growth (Alexander et al., 2005), typically only a few percent show evidence of harbouring a highly-accreting quasar (Almaini et al., 2003). 

Studies of obscured (or X-ray absorbed; N$_{H}$~$>$~10$^{22}$~cm$^{-2}$) quasars show evidence that many accreting active galactic nuclei (AGN) are hidden within kpc-scale dust envelopes (Page et al., 2004; Martinez-Sans\'{i}gre et al., 2009). This is consistent with an evolutionary picture whereby a period of AGN growth occurs within young, dusty, gas-rich galaxies. However if the end-point in the evolution of massive galaxies is a relaxed spheroid containing a supermassive black hole then, given the rarity of X-ray-bright sub-mm galaxies, the most intense periods of growth of these two principal components does not appear to be synchronised over the lifetime of the galaxy.

In order to match observations, most models of galaxy formation require a period of AGN activity, the resulting outflows eventually suppressing star formation (a mechanism known as feedback; e.g.~Silk \& Rees, 1998). Models that invoke the starburst and quasar phase as part of an evolutionary sequence (Archibald et al., 2002) have been proposed, whereby the quasar phase of the AGN occurs only after the black hole has had time to grow sufficiently ($\sim$0.5~Gyr). The quasar activity thus lags the bulk star formation event, which may itself be triggered earlier in the lifetime of the galaxy by an event such as a major merger. Massive outflows from a quasar have recently been directly observed in the early Universe ($z$~=~6.4189; Maiolino et al., 2012), observational evidence that such activity directly suppresses star formation remains elusive (Harrison et al., 2012).

The radio loud quasar 3C318 (RA~=~15h~20m~05.49s, Dec~=~+20$^{\circ}$~16'~05.71", J2000) represents an excellent case study for examining the interplay between star formation and the jets and radiation emitted by AGN within the epoch of peak activity for both of these phenomena. It is a compact steep spectrum source, and thus is a young radio source (Snellen, 2008) with jets that were triggered relatively recently. Long baseline radio observations with the VLBA and MERLIN corroborate this scenario, resolving asymmetric jets with a maximal extent of 0.8 arcseconds ($\sim$7~kpc in projection) in a south-westerly direction (Spencer et al., 1991; Mantovani et al., 2010). Assuming the jets advance at a speed of 0.1c their age is $<$1~Myr (Willott et al., 2007).

There are strong infrared \emph{IRAS} (Hes et al., 1995) and sub-mm SCUBA (Willott et al., 2002) detections of 3C318. Willott et al.~(2000) spectroscopically measured a redshift of 1.574 from the quasar spectrum. Its hyperluminous properties at infrared wavelengths require very large dust mass, with quasar heating dominating the far infrared flux and contributing at a $\sim$10\% level to the sub-mm flux (Willott et al., 2000).

In order to better constrain the implied high stellar contribution to the dust heating, Willott et al. (2007) initiated molecular line observations using the Plateau de Bure Interferometer (PdBI). Carbon monoxide is the second most abundant molecule in galaxies (after H$_{2}$). It is the best observational tracer as the $J$~=~1~$\rightarrow$~0 ($\nu_{rest}$~=~115.27~GHz) transition of the $^{12}$C$^{16}$O (hereafter CO) molecule has a readily achievable excitation temperature of 5.5~K. Although the CO line is not a good indicator of the star-formation rate it offers a powerful tool for studying the physical conditions of the gas which makes up the star formation budget of a galaxy (Solomon \& Vanden Bout, 2005). Observations of the $J$~=~2~$\rightarrow$~1 transition of CO in 3C318 inferred a H$_{2}$ mass of 3.0~($\pm$0.6)~$\times$~10$^{10}$~($\alpha_{CO}/0.8$)~M$_{\odot}$, comparable to that of a typical sub-mm galaxy at a similar redshift.

This article presents high angular resolution Very Large Array (VLA) observations of 3C318 at 44~GHz, targeting the redshifted CO $J$~=~1~$\rightarrow$~0 transition. The primary goal is to spatially resolve the system and determine whether the molecular gas lies within the quasar host galaxy suggesting a circumnuclear starburst, or is spatially offset from the radio core, as would be expected from an on-going major merger scenario. Note that there is no evidence for strong gravitational lensing effects in 3C318 (Willott et al., 2000).

A description of the data acquisition and reduction is presented in Section \ref{sec:observations}. The continuum and continuum-subtracted line maps are presented and these results are discussed in Section \ref{sec:results}. Section \ref{sec:conclusions} contains concluding remarks. Throughout this paper the following cosmological parameters are assumed: $H_{0}$~=~71 km~s$^{-1}$~Mpc$^{-1}$, $\Omega_{M}$~=~0.27, $\Omega_{\Lambda}$~=~0.73 (Komatsu et al., 2011).

\section{Observations and data reduction}
\label{sec:observations}

\begin{figure}
\nonumber
\centering
\includegraphics[width= \columnwidth]{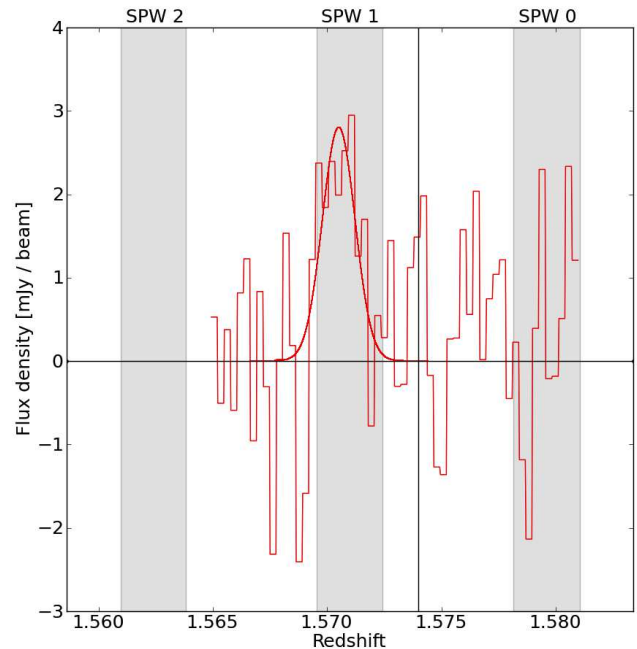}
\caption{The histogram shows the CO $J$~=~2~$\rightarrow$~1 spectrum from the PdBI observations of Willott et al.~(2007), overlaid with the fitted Gaussian profile. The three vertical bands show the spectral coverage of the VLA observations presented in this paper. The vertical line shows the redshift of the quasar as derived from the narrow emission lines (Willott et al., 2000). A fraction of the emission from the Gaussian line profile is missed by the limited frequency coverage of the VLA observations. A correction factor for this is derived in Section \ref{sec:lineproperties}.
\label{fig:pdb_spec}}  
\end{figure}

3C318 was observed with the VLA in C-configuration in four 5.75 hour sessions between 22 December 2006 and 11 January 2007. The array consisted of 26 antennas, 7 of which were fitted with new Expanded Very Large Array (EVLA) receivers.

A moderately fast switching cycle was employed, switching between 3C318 and the nearby phase calibrator source J15169+19322 with a 150:40-second duty cycle. Flux scaling was achieved by observing 3C286 for five minutes during each of the four sessions.

The old VLA correlator had not yet been replaced, and the spectral setup consisted of three spectral windows (SPWs; a.k.a.~IFs) at 44.6851, 44.8351 and 44.9851 GHz, with a bandwidth of 50~MHz, recording all four polarization products. As can be seen in Figure \ref{fig:pdb_spec}, the central SPW1 at 44.8351 would thus contain the redshifted 115.27 GHz CO $J$~=~1~$\rightarrow$~0 line for 3C318 at $z$~=~1.5705, with the assumed redshift of the target being that of the aforementioned PdBI detection of the CO $J$~=~2~$\rightarrow$~1 line (Willott et al., 2007), offset from the systemic redshift (1.574) derived from the narrow emission lines by -400~km~s$^{-1}$. Note that the $J$~=~1~$\rightarrow$~0 transition at the quasar redshift falls into a gap betweeen two spectral windows. 

The other two SPWs were set up to record the continuum on either side of the line. Only two SPWs were recorded simultaneously, thus the line was targeted continuously in SPW1 and the two continuum frequencies were switched approximately every 25 minutes. 

Data were edited, calibrated and imaged using the {\tt CASA}\footnote{\tt http://casa.nrao.edu} package. The data were initially inspected and flagged using the {\tt PlotMS} tool. Complex antenna gain solutions were derived for the phase calibrator which were then interpolated in time across the target source. The primary flux calibrator 3C286 was only observed in SPWs 0 and 1 so the solutions were extrapolated for SPW 2 when flux density scaling was applied. Examining the integrated flux densities in Table \ref{tab:fluxes} demonstrates the accuracy of this process, with the measured flux densities of the phase calibrator having a spread of 1~$\mu$Jy, or 0.1\% of the mean value.
\begin{table}
\centering
\begin{tabular}{cccc} \hline
& & \multicolumn{2}{c}{Flux density (mJy)} \\
SPW & Frequency (GHz) & J1516+1932 & 3C318 \\ \hline
0 & 44.6851 & 1.0024 ($\pm$0.0005) & 93.0 ($\pm$1.5) \\
1 & 44.8351 & 1.0016 ($\pm$0.0004) & 92.4 ($\pm$1.5) \\ 
2 & 44.9851 & 1.0026 ($\pm$0.0006) & 91.5 ($\pm$1.6) \\ \hline
\end{tabular}
\caption{Frequency of the three spectral windows, and the integrated flux densities and associated
uncertainties derived by fitting two-dimensional Gaussians to the emission in naturally-weighted maps
of both 3C318 and the phase calibrator using the {\tt AIPS} task {\tt IMFIT}. Note that the uncertainties
quoted here are those of the integrated flux of the fitted Gaussians, and do not correspond to the map noise.\label{tab:fluxes}}
\end{table}
%
	
\section{Results and discussion}
\label{sec:results}


\begin{figure*}
\begin{center}
\setlength{\unitlength}{1cm}
\begin{picture}(16,6)
\put(-1.45,-1.0){\includegraphics{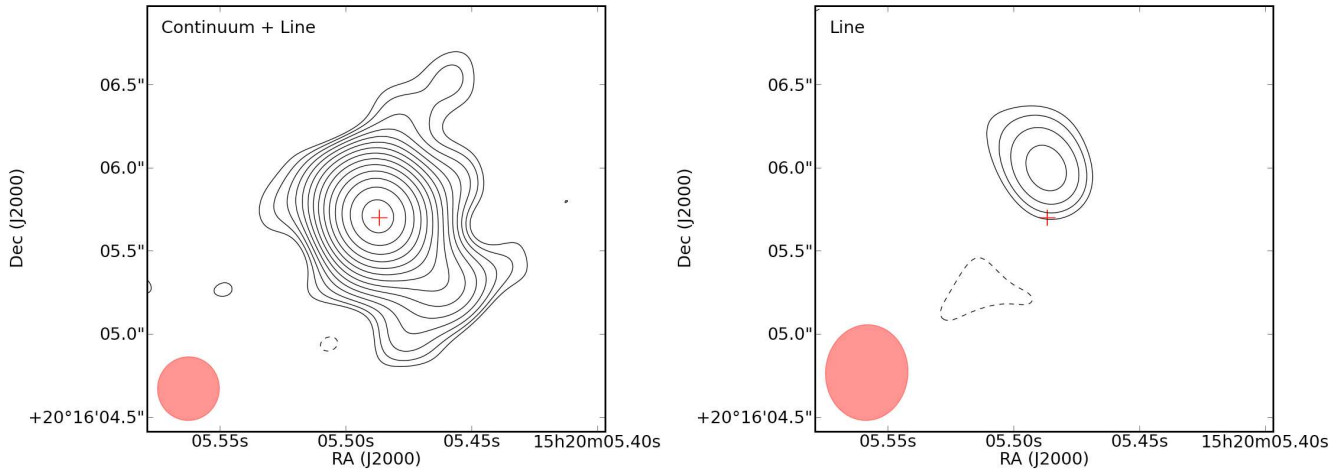}}
\end{picture}
\caption{Contour images of 3C318 at 44.8 GHz. The left hand panel shows the continuum + line image and the right hand panel shows the continuum-subtracted line emitter. Contour levels are (--1, 1, $\sqrt{2}$, 2, 2$\sqrt{2}$, 4, 4$\sqrt{2}$, 8, 8$\sqrt{2}$...)~$\times$~3$\sigma$, where $\sigma$ is the RMS background noise in the image. For the continuum + line image $\sigma$~=~0.18~mJy per beam, and for the line map $\sigma$~=~0.148~mJy per beam. The differing noise values are due to the weighting schemes used during imaging. The left hand image is generated using Briggs (robust~=~0) weighting to better display the radio jet, whereas the right hand image is generated using natural weighting. The extent of the restoring beam is shown in the ellipses in the lower-left corner of each map. The restoring beams are 0.38~$\times$~0.35 arcseconds (position angle~=~2 degrees west of north) and 0.58$\times$~0.46 arcseconds (position angle~=~2.8 degrees west of north) for the left and right panels respectively. The cross in the centre of each image shows the position of the radio peak.
\label{fig:radio_maps}}  
\end{center}
\end{figure*}

\subsection{Continuum image}
\label{sec:continuum}

A Briggs (1995) weighted (robust~=~0) map of the continuum + line emission generated by combining data from all three spectral windows is shown in the left hand panel of Figure \ref{fig:radio_maps}. The root-mean-square (RMS) background in this image is $\sigma$~=~0.18~mJy per beam and the contour levels are provided in the figure caption. The radio jet is partially resolved at 115 GHz rest-frame. The angular distance from the peak of the emission coincident with the core, along the radio jet to to the lowest contour in the map is 0.74 arcseconds, corresponding to a projected linear size of 6.33~kpc, consistent with the 5 GHz VLBA (Mantovani et al., 2010) and 1.6 GHz MERLIN (Spencer et al., 1991) observations.

\subsection{CO $J$~=~1~$\rightarrow$~0 detection}
\label{sec:co}

The molecular line emission was searched for in the image-plane by producing and deconvolving a naturally-weighted continuum image from the visibilities in the two spectral windows which did not cover the CO $J$~=~1~$\rightarrow$~0 line.  A second naturally-weighted, deconvolved `line~+~continuum' image was then formed from the visibilities in the single spectral window (SPW1) that covered the line. The former was subtracted from the latter, and the residual image resulting from this process is shown in the right hand panel of Figure \ref{fig:radio_maps}. An unresolved source with peak flux density of 1.52 mJy per beam is present. The RMS value of the background noise in this map is 0.148 mJy per beam, thus the residual feature representing the CO emitter has a signal to noise ratio of $\sim$10. The RMS noise in the line map ($\sigma$~=~0.148 mJy per beam) is taken to be the error in the measured flux density, and is lower than that of the continuum map ($\sigma$~=~0.18~mJy per beam) due to the different weighting schemes employed, and also since the residual image is derived from the subtraction of two images both containing noise of similar levels. In theory this should increase the noise in the map by a factor of $\sqrt{2}$. Contour levels are again provided in the figure caption.

To ensure that the detection of the line emitter is not a spurious noise peak the data were split into two chunks, each containing half of the scans, re-imaged and re-subtracted. The line source persisted for both of these split data sets, albeit with a correspondingly lower signal to noise ratio. Alternative residual images were also generated by subtracting the maps derived from individual continuum spectral windows from a line image formed from appropriately paired scans in order to match the sensitivity. In each case the source persisted. Subtraction of the two continuum images formed from SPWs 0 and 2 resulted in noise-like residuals. The line source also remained when the data were imaged with different imaging weights. Natural weighting was chosen in the final image to maximise sensitivity. These tests argue against the line emitter being an imaging artefact such as a sidelobe, as both the splitting of the data (in both time and frequency) and changing the imaging weights results in different point spread functions.

\subsection{The spatial offset of the molecular line emitter}
\label{sec:offset}

\begin{figure}
\nonumber
\centering
\includegraphics[width= \columnwidth]{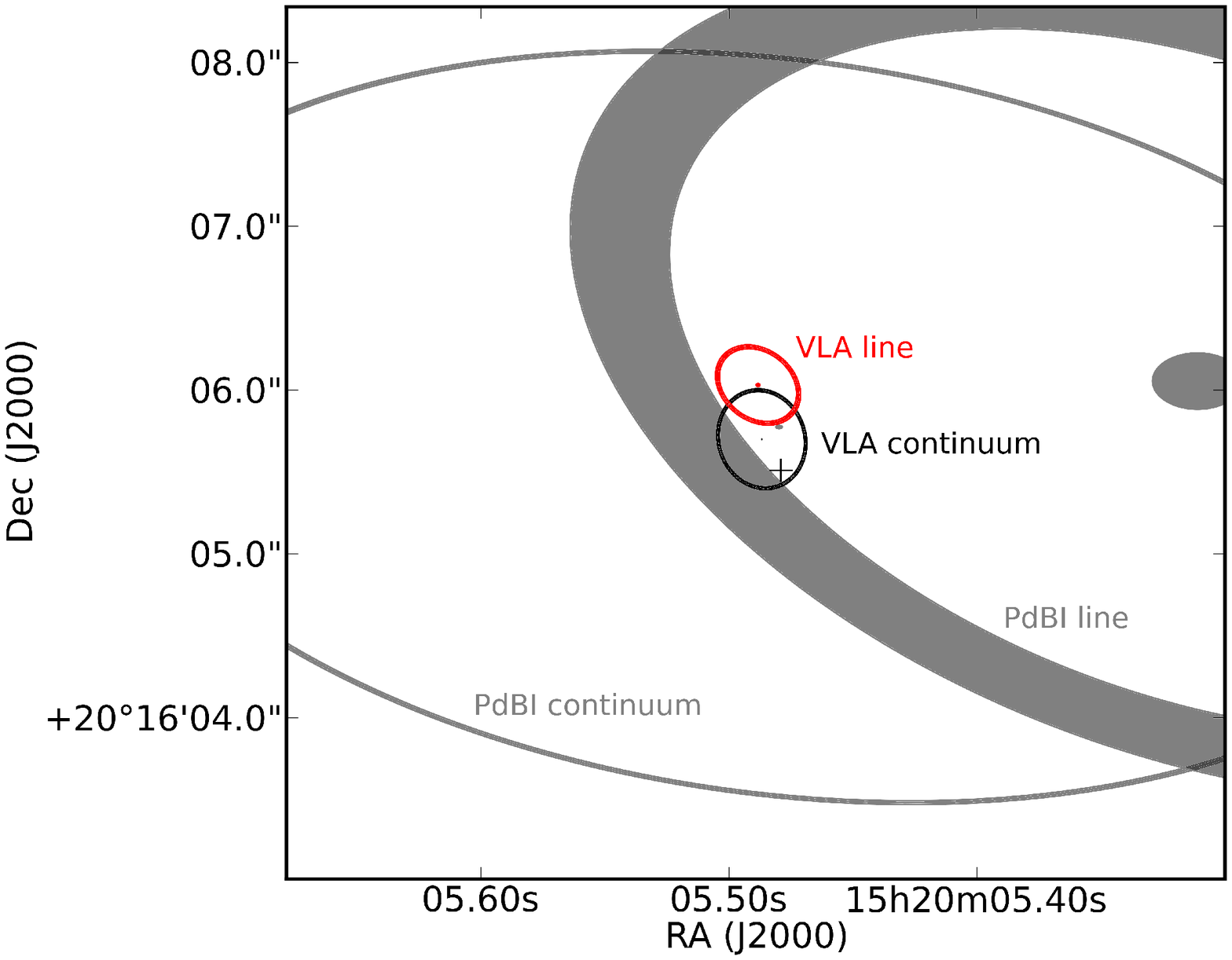}
\caption{Visualisation of the data in Table \ref{tab:gaussians}; properties of two-dimensional Gaussian components fitted to the molecular line emitter and the continuum peak for both the VLA and the PdBI observations. The large ellipses show the major and minor axes of the fitted Gaussians and the smaller concentric ellipses show the corresponding fitted positions. The thickness of the large ellipses shows the $\pm$1$\sigma$ uncertainties in the axial fits and the size of the central ellipse shows the $\pm$1$\sigma$ positional uncertainty. The high signal-to-noise ratio detections afforded by the high resolution VLA observations result in extremely tight positional constraints on the fitted Gaussians, which in turn demonstrates the high significance ($>$8$\sigma$) of the spatial offset of the line emitter in 3C318. The SDSS position of the optical quasar is shown by the plus sign. The uncertainties in this position are not plotted, but they are quoted as 0.5 arcseconds in both RA and Dec, i.e.~consistent with the radio peaks within those margins.}
\label{fig:ellipses}
\end{figure}

The crosses in both panels of Figure \ref{fig:radio_maps} mark the position of the peak of the continuum emission, highlighting the fact that the line emitter is spatially offset from the nucleus of 3C318. To assess the significance of the observed separation, a single two-dimensional Gaussian is fitted to the emission from the line emitter and the continuum peak for both the VLA and the PdBI observations. Contrary to the continuum + line VLA map shown in Figure \ref{fig:radio_maps} the Gaussian fitting is performed on a naturally weighted map in order to match the resolution of the line-only image. Fits were peformed using the {\tt IMFIT} task within the {\tt AIPS} package (Greisen, 2003). The returned parameters are listed in Table \ref{tab:gaussians}, and visualised in Figure \ref{fig:ellipses}, together with the associated uncertainties. With the exception of the spatially resolved (even with natural weighting) VLA continuum + line image, the fitted extents of the Gaussians in the images were all consistent with the restoring beams to within their uncertainties. 

Also present in Table \ref{tab:gaussians} are the estimates of the intrinsic source sizes as determined by the deconvolution of the restoring beam. In the case of the spatially extended VLA continuum + line image the fit was core dominated such that the deconvolved major axis size was significantly less than the actual extent of the radio jet, as discussed in Section \ref{sec:continuum}.  As such the estimated intrinsic extent of this component was considered unreliable, and is omitted from Table \ref{tab:gaussians}. The only parameter returned by the deconvolution process that will feature in the subsequent analysis is the upper limit to the extent of the line emission region as measured from the VLA image (Section \ref{sec:lineproperties}).

The calculated offset between the peak of the continuum emission and the line emitter, using the fitted parameters of the VLA observations and propagating their uncertainties, is 0.33 ($\pm$0.02) arcseconds. The offset, \emph{considering only the uncertainties in the fitted positions} (see below), is thus significant at the $\sim$17$\sigma$ level. At a redshift of 1.574 using the assumed cosmology this corresponds to a physical distance of 2.82~kpc projected onto the plane of the sky. The high resolution and high significance of the detection in the VLA observations provides a substantial refinement of the position of the line emitter and its separation previously reported by Willott et al. (2007). The offset in the PdBI observations is also presented in Table \ref{tab:gaussians}, which corresponds to a physical separation of 21.8~kpc\footnote{This is consistent with the $\sim$20~kpc value reported by Willott et al.~(2007) when the updated cosmological parameters are taken into account.}. 

Immediately obvious from the calculation of the offsets for the VLA and PdBI data is that the two derived values do not overlap to within the errors (although the VLA positions are within the FWHM of the PdBI restoring beam as can be seen in Figure \ref{fig:ellipses}). How can this be explained, starting with the assumption that there is no physical explanation for an intrinsic spatial offset of this significance between the two lowest $J$ transitions of a CO line emitter? Note that the errors in the parameters of a two-dimensional Gaussian fit to a radio source are generally smaller than the true uncertainties. The positional uncertainty of a point source is a combination of astrometric error (usually induced by calibration errors) and a statistical uncertainty which is dependent on the signal to noise ratio of the detection and the resolution of the instrument (Condon, 1997). Fitting a Gaussian is a process that is sensitive to only the statistical component.

In determining the level of the calibration induced offsets, radio surveys generally make use of external data sets, for example calibrator lists with accurate positions derived from Very Long Baseline Interferometry. Many sources have their offsets in Right Ascension and Declination measured with respect to the reference sources, and these measurements are subsequently used to correct any systematic mispointing and provide a measure of the scatter in the measured source positions. Observations at L-band typically result in offsets in both Right Ascension and Declination with a standard deviation of $\sim$0.1" (Prandoni et al, 2000; Bondi et al., 2003). Assuming this scatter scales linearly with frequency, one would expect the corresponding value at 44~GHz to be $\sim$3.2 milliarcseconds. 

Although the VLA observations in this paper do not contain large numbers of sources to perform such a check, a rough handle on the positional scatter in the VLA observations of 3C318 can be obtained by comparing the measured positions of the phase calibrator with previously published positions at high resolution. Imaging the phase calibrator using data from each of the three separate days over which these observations were taken and fitting a Gaussian yields variations in the fitted source position of $\sim$1 milliarcsecond in both Right Ascension and Declination. This is consistent with the scaled L-band estimate provided above. The measured positions of the phase calibrator are also consistent with the position derived from an 8.4 GHz VLA calibrator search (Browne et al., 1998) to within $\sim$1 milliarcsecond.

The estimated astrometry errors induced by calibration thus appear to be present in the VLA data at a level that is about 15\% of the statistical uncertainties in the Gaussian fit to the line emitter. Including an additional 3 milliarcsecond contribution to the uncertainty in the two VLA positions used for calculating the offset increases the error in the offset measurement to $\pm$0.04", and reduces the significance of the offset to 8.3$\sigma$.

Even with this increased contribution to the positional uncertainties in the VLA measurements the offset is still inconsistent with the value derived from the PdBI imaging. The cause of this remains unknown. It is worth mentioning that the statistical errors induced in the PdBI position measurements are much greater than those of the VLA measurements due to the significantly lower spatial resolution and signal to noise ratio of the detections. This almost certainly plays a part, but there may also be some residual calibration error in the PdBI data that cannot be quantified here. Note also that even the fitted positions of the strong quasar peak which is detected at high significance by both instruments differs by approximately 0.1", significantly higher than the assumed VLA systematic uncertainties. Further high frequency observations of 3C318 with higher sensitivity should unravel this issue.

\begin{table*}
\begin{minipage}{178mm}
\centering
\begin{tabular}{lcccc} \hline
               & \multicolumn{2}{c}{VLA}       & \multicolumn{2}{c}{PdBI} \\
               & \multicolumn{1}{c}{Continuum + Line} & \multicolumn{1}{c}{Line}       & \multicolumn{1}{c}{Continuum + Line}  & \multicolumn{1}{c}{Line}\\ \hline
\multicolumn{5}{l}{\emph{Positional fits}} \\
Right Ascension             & 15h 20m 5.4868s ($\pm$0.0005")               & 15h 20m 5.49s ($\pm$0.02")                 & 15h 20m 5.48s ($\pm$0.04")                 & 15h 20m 5.3s ($\pm$0.5")\\ 
Declination            & 20$^{\circ}$ 16' 5.6996" ($\pm$0.0006")      & 20$^{\circ}$ 16' 6.03" ($\pm$0.02")        & 20$^{\circ}$ 16' 5.77" ($\pm$0.02")        & 20$^{\circ}$ 16' 6.1" ($\pm$0.3")\\ 
\multicolumn{5}{l}{} \\
\emph{Line/continuum offset} & \multicolumn{2}{c}{{\bf 0.33 ($\pm$0.02)" (2.82~kpc)}} & \multicolumn{2}{c}{{\bf 2.54 ($\pm$0.51)" (21.8~kpc)}} \\ 
\multicolumn{5}{l}{} \\
\multicolumn{5}{l}{\emph{Image plane Gaussian fits}} \\
Fitted maj.~axis     & 0.62 ($\pm$0.03)"             & 0.54 ($\pm$0.05)"              & 8.18 ($\pm$0.08)"              & 7.51 ($\pm$1.19)"\\
Fitted min.~axis     & 0.50 ($\pm$0.03)"             & 0.42 ($\pm$0.04)"              & 4.50 ($\pm$0.04)"              & 4.52 ($\pm$0.72)"\\
Fitted PA & 12.1 ($\pm$0.4)$^{\circ}$   & 43.3 ($\pm$14.9)$^{\circ}$   & 81.4 ($\pm$0.6)$^{\circ}$    & 66.6 ($\pm$12.1)$^{\circ}$\\ 
\multicolumn{5}{l}{} \\
\multicolumn{5}{l}{\emph{Deconvolved source properties}} \\
Major axis     & --            & $R_{maj}$~$<$~0.23"              &   $R_{maj}$~$<$~2.0"            & 2.45"~$<$~$R_{maj}$~$<$~3.73" \\
Minor axis     & --            & Unresolved              &   $R_{min}$~$<$~2.0"           & Unresolved \\
PA                   & -- & 8.5$^{\circ}$~$<$~PA~$<$~57.6$^{\circ}$   & 33.9$^{\circ}$~$<$~PA~$<$~47.2$^{\circ}$    & 70.5$^{\circ}$~$<$~PA~$<$~144$^{\circ}$\\ \hline

\end{tabular}
\caption{Positions and extents of the two-dimensional Gaussians fitted to the line and continuum peaks for both the VLA observations presented in this paper and the PdBI observations of Willott et al. (2007), together with the associated 1$\sigma$ uncertainties in the fitted parameters. Estimates of the intrinsic source sizes are are also presented following deconvolution of the restoring beam. Note that the positional offsets between the line and the continuum have been calculated using only the positional fits and their associated uncertainties quoted here. The positional errors associated with Gaussian fits are generally smaller than the true positional errors in the observation which have a systematic contribution generally imposed by residual calibration errors. See the text for full details.\label{tab:gaussians}}
\end{minipage}
\end{table*}

Despite the astrometric uncertainties outlined above, the bulk of the molecular gas associated with 3C318 is clearly significantly offset from the quasar nucleus. How should this offset be interpreted? It is notable that in the sky projection a line from the quasar nucleus to the line emitter is offset from the jet axis by $\simeq$220$^{\circ}$ west of north, or 40$^{\circ}$ east of north for an undetected counterjet assumed to lie on a common axis. Such an offset rules out jet-induced star formation (De Young, 1989) as the cause of the observed star formation activity in 3C318. Given that the observed molecular emission in 3C318 is not co-spatial with the quasar nucleus there is little evidence for a circumnuclear starburst being responsible for the ultraluminous infrared emission. 

A strong possibility is that 3C318 is undergoing a major merger. Molecular line observations can directly probe these events when the quasar host galaxy and the companion galaxy are gas rich, and such observations have been presented by several authors. For example line and continuum observations of the $z$~=~4.7 quasar BR1202-0725 (Omont et al., 1996; Carilli et al., 2002) are spatially resolved into two components corresponding to a luminous quasar host galaxy and a bright sub-mm galaxy with a projected separation of 25~kpc. The CO $J$~=~5~$\rightarrow$~4 lines associated with each component exhibit velocity offsets, arguing against gravitational lensing being responsible for the multiple imaging. Recent ALMA observations of this system at 345~GHz detect thermal dust continuum and [CII] lines in both components, supporting the merger scenario (Wagg et al., 2012). Similarly, EVLA observations of the $z$~=~6.18 quasar J1429+5447 reveals two peaks in CO $J$~=~2~$\rightarrow$1 with a separation of 6.9 kpc, one of which is coincident with the quasar position (Wang et al., 2011). High resolution imaging of the quasar BRI 1335-0417 ($z$~=~4.41; Riechers et al., 2008) reveals evidence of a major merger in the form of multiple components extended over $\sim$5~kpc around the radio quiet (Momjian et al., 2007) quasar nucleus with an overall velocity spread of $\sim$300~km~s$^{-1}$, thus on a very similar scale to 3C318. The molecular gas peaks are embedded in a structure which is inconsistent with a rotating disk, exhibiting instead the complex velocity structure expected from a highly-disrupted system. The single peak in the line image in Figure \ref{fig:radio_maps} and the Gaussian-like line profile of the CO $J$~=~2~$\rightarrow$~1 PdBI observations suggest that the velocity structure of 3C318 is somewhat more orderly, although multiple velocity components on scales not probed by the current observational limits cannot be ruled out.

Another possibility is that 3C318 contains a circumnuclear ring of molecular material. The nearby radio-loud Fanaroff-Riley Type-I (FR-I; Fanaroff \& Riley, 1974) galaxy NGC 3801 contains 3~$\times$~10$^{8}$~M$_{\odot}$ of H$_{2}$ in a rotating ring, inclined at 84$^{\circ}$ to the axis of the radio jets (Das et al., 2005). The radius of this ring is $\sim$2~kpc, thus at a similar separation from the nucleus as the line emitter seen in 3C318, and NGC 3801 is known to be a merger remnant with young radio jets (Hota et al., 2012). The powerful FR-II radio galaxy 3C293 is also known to contain a large reservoir (M$_{H_{2}}$~=~1.5~$\times$~10$^{10}$~M$_{\odot}$) of molecular gas distributed in an asymmetric ring within the central 3~kpc (Evans et al., 1999). Again, a previous merger event is invoked in order to explain the gas ring and the triggering of the radio jets. If 3C318 has a similar ring of molecular material then the VLA image above is likely to only show the blueshifted material due to the limited spectral coverage, however the PdBI spectrum would have enough bandwidth to detect a corresponding redshifted component if it were there (Figure \ref{fig:pdb_spec}). The evolved systems cited above with molecular gas rings could represent the final phase of the merger scenario, with 3C318 representing and initial stages and BRI 1335-0417 being observed at an intermediate time, exhibiting the most dynamically perturbed velocity structure.

\subsection{Properties of the molecular line emitter}
\label{sec:lineproperties}

It is assumed that the CO $J$~=~1~$\rightarrow$~0 emission is distributed over a Gaussian line profile with the same 200~km~s$^{-1}$ FWHM as the CO $J$~=~2~$\rightarrow$~1 emission line that was spectroscopically resolved by Willott et al.~(2007). Integrating the Gaussian profile shown in Figure \ref{fig:pdb_spec} between the limits of SPW 1 shows that the VLA observations are missing 9\% of the emission in the wings of the line. Factoring this in results in a velocity-integrated line flux of the CO $J$~=~1~$\rightarrow$~0 line of 0.353~($\pm$0.034)~Jy~km~s$^{-1}$. Following Solomon and vanden Bout (2005), the CO line luminosity in K~km~s$^{-1}$~pc$^{2}$ can be determined by
\begin{equation}
L'_{CO}~=~3.25\times10^{7}~I_{CO}~\nu_{obs}^{-2}~D_{L}^{2}~(1+z)^{-3}
\end{equation}
where $I_{CO}$ is the velocity-integrated CO line flux in Jy~km~s$^{-1}$, $\nu_{obs}$ is the observing frequency in GHz, $D_{L}$ is the luminosity distance in Mpc and $z$ is the redshift of the source. At $z$~=~1.574 for the assumed cosmological parameters the luminosity distance is 11.689 Gpc, resulting in a value for $L'_{CO}$ of 4.6~($\pm$0.5)~$\times$~10$^{10}$~K~km~s$^{-1}$~pc$^{2}$. The corresponding molecular gas mass is 3.7~($\pm$0.4)~$\times$~10$^{10}$~M$_{\odot}$, assuming
\begin{equation}
M_{H_{2}}~=~\alpha_{CO}~L'_{CO}
\end{equation}
where $\alpha_{CO}$~=~0.8~M$_{\odot}$~(K~km~s$^{-1}$~pc$^{2}$)$^{-1}$. This is the value derived for ultra luminous infrared galaxies (ULIRGs; Downes and Solomon, 1998) and is typically employed when deriving gas masses for high-$z$ sub-mm galaxies and quasars (e.g.~Carilli et al., 2010; Riechers, 2011a). Empirical values for $\alpha_{CO}$ have a large range, with the ULIRG value at one end and the value measured from giant molecular clouds in the Milky Way at the other ($\alpha_{CO}$~$\simeq$~5; Solomon et al., 1987). Discussions of the origin and implications of this large uncertainty are presented by e.g.~ Obreschkow et al. (2009b, Appendix B) and Ivison et al. (2011). The value of 0.8 adopted here facilitates comparison with other similar observationally derived values. 

Following Neri et al.~(2003) the dynamical mass of a rotating disk can be estimated via
\begin{equation}	
M_{dyn} \sin^{2}(i) = 4 \times 10^{4} \Delta v^{2}R
\end{equation}
where $M_{dyn}$ is the dynamical mass in M$_{\odot}$, $i$ is the inclination angle, $\Delta v$ is the FWHM of the line width in km~s$^{-1}$ and $R$ is the outer radius of the disk in kpc.

Adopting for the value of $R$ the upper limit of the resolved axis of the beam-deconvolved line emitter (Section \ref{sec:offset}, Table \ref{tab:gaussians}) the spatial extent of the disk is 1.97 kpc. The corresponding limit on the inclination-dependent dynamical mass is therefore $M_{dyn}$~sin$^{2}$($i$)~$<$~3.2~$\times$~10$^{9}$~M$_{\odot}$. The ratio of the molecular gas to the dynamical mass $M_{H_{2}}$~/~$M_{dyn}$~$>$~12~($\alpha_{CO}$/0.8)~sin$^{2}$($i$), so M$_{H_{2}}$ exceeds M$_{dyn}$ if $i$~$>$~16$^{\circ}$ if $\alpha_{CO}$~=~0.8. Observations of high-$z$ galaxies which measure both the dynamical and molecular gas masses show a range of M$_{H_{2}}$/M$_{dyn}$ ratios. In the nearby radio loud galaxy NGC 3801 this fraction is $\sim$1\% (Das et al., 2005), in sub-mm galaxies Tacconi et al.~(2006; 2008) determine fractions in the range 20--60\% and Daddi et al.~(2010) find fractions as high as 65\% in normal star forming galaxies at $z$~=~1.5. Note also that uncertainties in the value of $\alpha_{CO}$ adopted for deriving the molecular gas mass in 3C318 will also propagate into this calculation, although increasing the value of $\alpha_{CO}$ towards higher values serves to lower the upper limit on the inclination angle. The upper limit for the inclination angle in 3C318 is thus derived by retaining $\alpha_{CO}$~=~0.8 and adopting a gas fraction of 0.6, consistent with the upper end of the distribution for sub-mm galaxies. This yields $i$~$<$~13$^{\circ}$.
	
The above assumes that the disk assumption is valid, and thus the value for the dynamical mass is almost certainly a lower limit in the case of 3C318. If the merger scenario is correct then the above calculation is based only on the molecular emission from the infalling companion galaxy and not from the quasar host galaxy, in which there is currently no evidence for significant quantities of molecular gas. Similarly if the system has been tidally disrupted, the sensitivity and spectral resolution of the existing observations are insufficient in order to probe the dynamics, which in turn are almost certainly not well-modelled by a dynamically stable rotating disk (e.g.~Riechers et al., 2008). 

A volume of gas in local thermodynamic equilibrium will exhibit a ratio of line temperatures ($T_{2\rightarrow1}$/$T_{1\rightarrow0}$) equal to unity if $k T_{e}$~$>>$~$h \nu_{J}$, where $k$ is Boltzmann's constant, $T_{e}$ is the excitation temperature, $h$ is Planck's constant and $\nu_{J}$ is the line frequency. In the case of the $J$~=~2~$\rightarrow$~1 transition this requires $T_{e}$~$>>$~11~K, a reasonable assumption. The line temperature ratios can be calculated for the molecular line emitter associated with 3C318,
\begin{equation}
r_{2\rightarrow1}~=~\left(\frac{I_{2\rightarrow1}}{I_{1\rightarrow0}}\right)\left(\frac{\nu_{1\rightarrow0}}{\nu_{2\rightarrow1}}\right)
\end{equation}
where $I_{n\rightarrow(n-1)}$ is the velocity integrated flux for the transition in question and $\nu_{n\rightarrow(n-1)}$ is the corresponding observation frequency. For 3C318 $r_{2\rightarrow1}$~=~0.84~($\pm$0.17).

Studies of the $J$~=~1~$\rightarrow$~0 transition of CO in a sample of high redshift (2.28~$<$~$z$~$<$4.69) quasars by Riechers et al.~(2006; 2011b) show that in all cases the CO excitation conditions are consistent with the bulk of the gas being thermalized up to the mid-level ($J_{upper}$~=~4) transitions. The mean value of $r_{3\rightarrow1}$ from Riechers et al. (2011b) is 0.99. There is no evidence for extended molecular gas reservoirs in their sample, and all CO spectral energy distributions (SEDs) are well fit by single component models of the gas. The same article contrasts the molecular gas content of the quasar host galaxies with observations of typical sub-mm galaxies at $z$~$\sim$~2, noting that the latter class of object often shows large quantities of low-excitation gas. It is noted that these line observations are consistent with quasars and sub-mm galaxies representing different stages in the lifetime of a massive galaxy.

The $r_{2\rightarrow1}$ ratio derived above for 3C318, while suggesting that the bulk of the gas probed by these two low-$J$ transitions is thermalized, may also hint at sub-thermal excitation in an appreciable fraction of the molecular gas reservoir. Since 3C318 appears to harbour both an active nucleus and a massive starburst, it may represent one of the rare class `overlap' objects hosting both of these phenomena within the models whereby quasars and sub-mm galaxies exist on an evolutionary timeline. 

\subsection{3C318 and the L$_{FIR}$-L'$_{CO}$ relation}

\begin{figure*}
\begin{center}
\setlength{\unitlength}{1cm}
\begin{picture}(14,14.5)
\put(-1.2,-0.6){\includegraphics{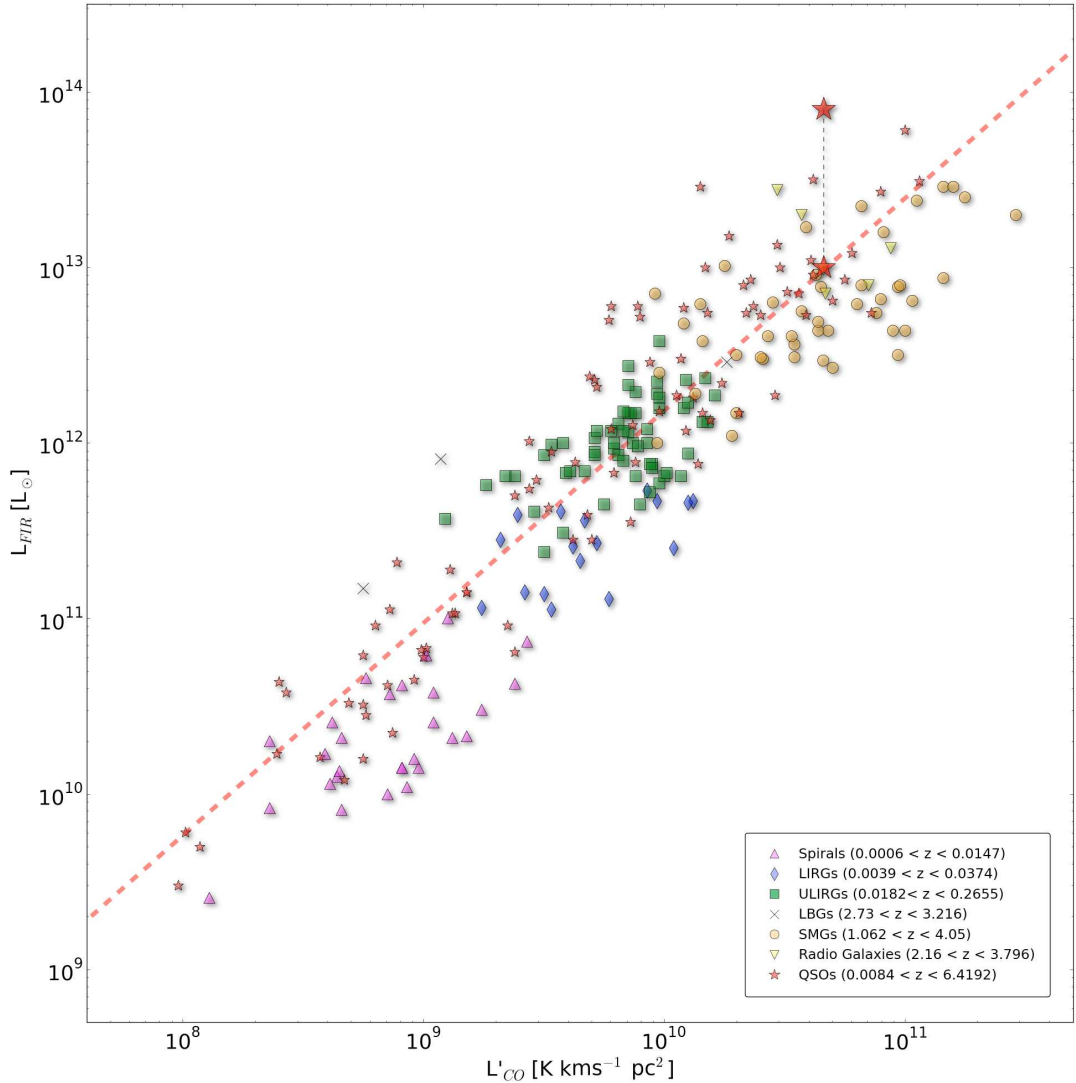}}
\end{picture}
\caption{Total far infrared luminosity against total CO line luminosity for 246 extragalactic systems collated from the literature plus 3C318. Spirals (Gao \& Solomon, 2004), (ultra-)luminous infrared galaxies ((U)LIRGs; Solomon et al., 1997; Gao \& Solomon, 2004; Chung et al., 2009), Lyman Break Galaxies (LBGs; Riechers et al., 2010; Magdis et al., 2012), sub-mm galaxies (SMGs; Solomon and vanden Bout, 2005; Carilli et al., 2010; Ivison et al., 2011; Harris et al., 2012; Thomson et al., 2012; Bothwell et al., 2013; Fu et al., 2013), radio galaxies (Solomon and vanden Bout, 2005; Emonts et al., 2013) and quasars (QSOs; Scoville et al., 2003; Walter et al., 2003; Solomon and vanden Bout, 2005; Evans et al., 2006; Riechers et al., 2006; Bertram et al., 2007; Maiolino et al., 2007; Aravena et al., 2008; Coppin et al., 2008; Riechers et al., 2008; Polletta et al., 2011; Riechers et al., 2011; Wang et al., 2011; Schumacher et al., 2012; Deane et al., 2013; Villar-Mart{\'{\i}}n et al., 2013; Xia et al., 2013; this paper) are included. Objects with only upper limits in either $L'_{CO}$ or $L_{FIR}$ are omitted. The symbols representing the various source classes are provided in the figure legend, along with the redshift ranges occupied by each class. 3C318 is indicated by the large star, and the lower value shows its FIR luminosity with the AGN contribution subtracted. The dashed line shows the Bothwell et al.~(2013) fit to the (U)LIRG and sub-mm populations. Intrinsic values corrected for lensing effects are plotted where applicable, and preference is given for values of $L'_{CO}$ determined from either observed or derived measurements of the intensity of the $J$~=~1~$\rightarrow$~0 transition. Further details are provided in the Appendix.}
\label{fig:lfir_lco}
\end{center}
\end{figure*}

Figure \ref{fig:lfir_lco} shows the far-infrared luminosity against the CO luminosity for 247 extragalactic systems (including 3C318) out to redshifts of $\sim$6.4. Spirals, (ultra-)luminous infrared galaxies (LIRGs and ULIRGs), Lyman Break Galaxies (LBGs), sub-mm galaxies (SMGs), radio galaxies and quasars (QSOs) are represented. References for each of these source types are provided in the figure caption, and the figure legend shows which symbol denotes which object class. The dashed line shows the Bothwell et al.~(2013) fit to the (U)LIRG and sub-mm populations.

Recall that the $x$ and $y$ axes on Figure \ref{fig:lfir_lco} are essentially proxies for the total molecular gas mass and the star formation rate respectively. Thus the ratio L$_{FIR}$/L'$_{CO}$ provides a measure of the gas depletion timescale (or star formation efficiency) of a system. In the case of 3C318 this is estimated to be 20~Myr, based on a FIR luminosity corrected for the AGN contribution to the dust heating (Willott et al., 2007). The large stars on Figure \ref{fig:lfir_lco} represent 3C318. The lower of the two corresponds to the FIR luminosity after the aforementioned correction has been applied, showing that 3C318 has a CO luminosity and thus a total molecular gas mass towards the upper end of the high redshift quasar population, and similar to that of a sub-mm galaxy. 

The systems that lie above the fitted trend are generally dominated by quasars, and the contribution of the AGN to the dust heating in the system is likely to be responsible for this. Correcting the FIR luminosity for 3C318 brings it in line with the fit. This is consistent with the FIR corrections in a sample of $z$~$\sim$~6 quasars presented by Wang et al.~(2011), and the picture presented by Willott et al.~(2007) that 3C318 hosts a starburst comparable to that of a typical sub-mm galaxy.

\section{Conclusions}
\label{sec:conclusions}

With a far-infrared luminosity typical of a highly-starforming sub-mm galaxy and a young, radio-loud active nucleus in quasar mode, 3C318 represents one of the best examples for studying the role of AGN and their associated jets on star formation in massive galaxies at the peak epoch of activity for both of these phenomena. 

Redshifted ground state emission from the CO molecule has been detected in 3C318 using the VLA at an observing frequency of 44~GHz. The system is spatially resolved and the molecular line emitter exhibits a projected 2.82~kpc positional offset from the quasar nucleus which is significant at the $>$8$\sigma$ level. Previously published spectroscopically resolved observations of the $J$~=~2~$\rightarrow$~1 CO line show that the line emitter is also offset by the systemic redshift of the quasar by -400~km~s$^{-1}$. 
The high resolution VLA images rule out a circumnuclear starburst or molecular gas ring, and suggest that the quasar host galaxy is likely to be undergoing a major merger with a gas rich galaxy, or that 3C318 is otherwise a highly disrupted system. The lack of significant line emission at the systemic redshift (from the PdBI observations) or position (from the VLA observations) of the quasar nucleus suggests that any merging galaxy is dominating the CO line emission for the two lowest $J$ transitions that have been observed in 3C318. Several high-$z$ quasars are seen to exhibit offsets between the core and the peak(s) of the molecular gas reservoirs, and such results are often interpreted as major merger events (Carilli et al., 2002; Riechers et al., 2008; Wang et al., 2011).

The 115 GHz rest-frame emission from the radio jet associated with the quasar is also resolved in the corresponding continuum map. As seen in existing long baseline observations at lower radio frequencies (Spencer et al., 1991; Mantovani et al., 2010), the jet of 3C318 has a small linear extent ($\sim$7~kpc), and was triggered $<$1~Myr ago (Willott et al., 2007). The position of the line emitter in relation to the jet axis also suggests that shocks from the radio jet are not inducing the observed high star formation rate in the system.

The $J$~=~1~$\rightarrow$~0 transition is the best tracer of the total amount of molecular gas in a system and the line luminosity of 4.6~($\pm$0.5)~$\times$~10$^{10}$~K~km~s$^{-1}$~pc$^{2}$ corresponds to a total H$_{2}$ mass of M$_{H_{2}}$~=~3.7~($\pm$0.4)~$\times$~10$^{10}$ M$_{\odot}$, assuming the typically-used luminosity-to-mass conversion factor of $\alpha_{CO}$~=~0.8~M$_{\odot}$~(K~km~s$^{-1}$~pc$^{2}$)$^{-1}$. If the line emitter can be modelled as a rotating disk then an inclination-dependent limit can be placed on the dynamical mass $M_{dyn}$~sin$^{2}$($i$)~$<$~3.2~$\times$~10$^{9}$~M$_{\odot}$, suggesting that M$_{H_{2}}$ is greater than M$_{dyn}$ if the inclination angle of this disk $i$~$>$~16$^{\circ}$. The derived CO line luminosity and thus H$_{2}$ mass are typical of luminous quasars and sub-mm galaxies at high redshift. 

Brightness temperature ratios between the CO line transitions measured to date suggest that the bulk of the molecular gas traced by these two lines in this system is in local thermodynamic equilibrium, although there is a hint of sub-thermal excitation of the $J$~=~2~$\rightarrow$~1 transition. Although CO observations of samples of high-$z$ quasars generally show thermalized emission up to the mid-$J$ transitions (Riechers et al., 2006; 2011b), additional higher temperature components in 3C318 cannot be ruled out on the basis of these observations alone. 

The two existing interferometric molecular line observations that have targeted 3C318 exhibit modest spatial resolution or modest spectral resolution but not both. Confirmation of the merger scenario could be achieved if high resolution line imaging could detect a second line peak at the quasar nucleus. Although the VLA following its upgrade now has sufficient spectral resolution and sensitivity to generate deep images of $J$~=~1~$\rightarrow$~0 emission in both the known line emitter and around the quasar nucleus, the next logical observation to make of 3C318 is to target higher $J$ CO line transitions using ALMA. Such an observation would allow modelling of the CO SED, constraining the physical conditions in the known line emitter. If a search for a circumnuclear reservoir of high-excitation gas being heated by the quasar revealed a second line peak, then such spectrally and spatially resolved observations would reveal the true dynamics of the system.

\section*{Acknowledgments}

We are very grateful to the anonymous referee for a diligent review of the manuscript, and comments that significantly improved it. I.H.~and A.~M.-S.~acknowledge the support of the South East Physics Network (SEPnet). A.~M.-S.~acknowledges a Post-Doctoral Fellowship from the United Kingdom Science and Technology Facilities Council, reference ST/G004420/1. The National Radio Astronomy Observatory is a facility of the National Science Foundation operated under cooperative agreement by Associated Universities, Inc. I.H.~thanks Hans-Rainer Kl\"{o}ckner and Aris Karastergiou for useful discussions. Some of the figures in this paper were produced with APLpy, an open-source astronomical plotting package for Python ({\tt http://aplpy.github.com}). This research has made use of NASA's Astrophysics Data System.

\appendix
\section{Collated L$_{FIR}$ and L'$_{CO}$ values for a range of extragalactic systems}

Figure \ref{fig:lfir_lco} shows the far-infrared and CO luminosities of 3C318 plus 246 extragalactic systems of various classes collated from the literature. Only solid detections are considered, and objects with only upper limits in either $L'_{CO}$ or $L_{FIR}$ are omitted from the sample. This collection is not fully exhaustive but it is representative. The publications that provided the original measurements are listed in the figure caption for each source class. Preference was given in all cases to luminosities derived from the $J$~=~1~$\rightarrow$~0 transition as this provides the best measurement of the total molecular gas mass (albeit one that is highly sensitive to the value of $\alpha_{CO}$). In some cases $J$~=~1~$\rightarrow$~0 luminosities were derived from higher $J$ transitions by the original authors using appropriately chosen correction factors. In this case the derived values are used in Figure \ref{fig:lfir_lco}. For publications where higher $J$ transitions were observed but $J$~=~1~$\rightarrow$~0 luminosities were not derived the correction factors listed by Carilli \& Walter (Table 2; 2013) were applied. Any known gravitational lensing corrections were also applied. The data for this table are available as supplementary material, or from the contact author.

\bsp 

\label{lastpage}

\end{document}